# Carrier Multiplication-Induced Structural Change during Ultrafast Carrier Relaxation and Non-Thermal Phase Transition in Semiconductors

Junhyeok Bang,[1,2] Y. Y. Sun,[1] X.-Q. Liu,[1] F. Gao,[3*] and S. B. Zhang[1*]

[1]*Department of Physics, Applied Physics, & Astronomy, Rensselaer Polytechnic Institute, Troy, NY 12180*

[2]*Spin Engineering Physics Team, Korea Basic Science Institute (KBSI), Daejeon 305-806, Republic of Korea*

[3]*Department of Nuclear Engineering and Radiological Sciences, University of Michigan, Ann Arbor, MI 48109*

*Corresponding authors: F. Gao (gaofeium@umich.edu) and S. B. Zhang (zhangs9@rpi.edu)

While being extensively studied as an important physical process to alter exciton population in nanostructures at fs time scale, carrier multiplication has not been considered seriously as a major mechanism for phase transition. Real-time time-dependent density functional theory study of $Ge_2Sb_2Te_5$ reveals that carrier multiplication can induce ultrafast phase transition in solid state despite that the lattice remains cold. The results also unify the experimental findings in other semiconductors for which the explanation remains to be the 30-year old phenomenological plasma annealing model.

PACS: 64.70.kg, 71.15.Mb, 71.15.Qe, 61.80.Az

Many physical processes may happen over a femtosecond (fs) time scale, ranging from simple atomic structural change to complex excited-state chemical reaction [1-6]. Non-thermal solid phase transition, involving massive atomic rearrangement, exemplifies such processes. Early experiments found that intense fs laser irradiation on semiconductors can lead to significant increase of reflectivity within sub-picoseconds (ps) [7-14]. Based on the assumption that the magnitude of the increase cannot be explained by any known direct effect of carrier excitation and the time scale of the change has exceeded the rate of energy transfer from electron to lattice, it was suggested that a structural transition has taken place non-thermally. Later, ultrafast-time-resolved X-ray diffraction experiments confirmed the suggestion and asserted that the non-thermal transition may be characterized by a sub-ps ionic inertial motion [15-18]. To explain the experimental findings, a plasma annealing (PA) picture was developed [19-21] in which the dense, excited carriers weaken the lattice to result in a shallow excited potential energy surface (PES) [as depicted in Figs. 1(a) and (b)].

Theoretical study of PA has been carried out extensively, assuming the excited electronic system is under a quasi-equilibrium condition [22-26]. As such, the excited carrier distribution may be described by the Fermi-Dirac distribution with a characteristic temperature and quasi-chemical potential. However, such an adiabatic approximation is often invalid in the time scale of sub-ps when the carrier-carrier (CC) and carrier-lattice (CL) scatterings [19,27,28], as well as the carrier dynamics, are all important. Figures 1(c) and (d) show two examples why this can be the case: carrier recombination by CL scattering [Fig. 1(c)] and carrier multiplication (CM) by CC scattering [Fig. 1(d)] [more examples can be found in Ref. 29]. A strong CL scattering rapidly increases the lattice temperature by energy transfer from the electrons to the lattice, which can be depicted as a vertical drop of the ionic PES in Fig. 1(a). This thermal activation of the lattice leads to phase transition by melting. In contrast, if CC scattering dominates, no significant energy transfer to lattice will take place, and the creation of secondary carriers moves the PES horizontally instead, as shown in Fig. 1(b). Whether a system undergoes a thermal or non-thermal phase transition in the sub-ps time regime thus depends critically on how the excitation energy is partitioned into the lattice and plasma branches. Note that CC scattering has been extensively studied for multi-exciton generation in semiconductor nanostructures for solar-cell applications [30] but its effects in solids have not been studied at the same level due to the lack of adequate first-principles approaches for excited-state dynamics.

Concerning the phase transition, recently Te-based phase-change memory (PCM) materials, e.g., the $Ge_2Sb_2Te_5$ (GST) alloys, have attracted considerable attention. Owing to the reversible transitions between crystalline and amorphous phases and the substantial changes in their electrical and optical properties, PCMs have been widely used for optical storage and are likely to be used for information technology should they be commercialized in non-volatile random-access memory [6,31-34]. Experiments on GST also revealed non-thermal phase transition [35,36], which can be highly desirable for improving device speed and reliability [37]. The experimental observations are also supported by first-principles calculations, but again under the steady-state approximation [38,39].

In this Letter, using GST as a prototype, we show that CM can lead to a switchover from predominantly thermal phase transition to non-thermal one. The use of a recently developed time-dependent density functional theory (TDDFT)–molecular dynamics (MD) method, which explicitly includes both the CC and CL scattering processes, enables us to explore the dynamic participation of the excitation energy during the early stage of the phase transition. We found that the magnitude of the excitation energy plays an essential role in determining the type of carrier relaxations: at low-energy (around the band gap) excitation (LEE), the CL scattering dominates. A significant temperature increase from the initial stage of the excitation is observed here, which aids the phase transition by heating. At high-energy (about 5 times the band gap) excitation (HEE), on the other hand, CM by CC scattering dominates, leading to a non-thermal phase transition significantly below the melting point, $T_m$ = 900 K [40]. These results remove the inconsistency between experiments [16,18] and theory (based on the PA model [26]). Non-thermal phase transition is not only important for GST but has also been reported experimentally for InSb, GaAs, Si, Se, and $VO_2$, [7-18,41,42]. Our theory could provide an understanding of the ultrafast carrier excitation/relaxation processes in other semiconductors as well.

Real-time coupled electronic and lattice dynamics are simulated based on TDDFT [43], as implemented in the code developed based on the SIESTA program [44-46]. Norm-conserving Troullier–Martins pseudopotentials [47], Perdew–Burke–Ernzerhof (PBE) exchange–correlation functional [48], and a local basis set with single-$\zeta$ polarized orbitals are employed. The real-space grid is equivalent to a plane-wave cutoff energy of 100 Ry. A supercell with 87 atoms (21 Ge, 18 Sb, and 48 Te) and 9 cation vacancies is employed [38]. $\Gamma$ point is used in the Brillouin zone integration. To mimic optical excitation by laser irradiation, we excite the electrons from the valence band to the conduction bands by changing their occupations [49,50]. This leaves holes in the valence band and excited electrons in the conduction band, as depicted in Fig. 2. In the dynamics calculations, we use a time step of 48 attoseconds and the Ehrenfest approximation for ion motion. An *NVE* ensemble is used to describe the effect of CL scattering. To prepare for the TDDFT-MD input structures, we perform electron-ground-state (GS) MD simulations to obtain equilibrated initial atomic coordinates and velocities at a lattice temperature of 670 K [38]. For an unbiased statistics, we consider ensembles in the MD - each consists of ten simulations with different initial atomic coordinates and velocities. To calculate the time evolution of the electron occupation in the adiabatic states, which are eigenstates of the Hamiltonian at any given time, we project out the time-evolved wave functions [29].

Figure 2 shows the density of states of GST in the crystalline phase. To discuss the effect of excitation, we consider a symmetric initial excitation of HEE as highlighted by the pink-colored regions, in which the electrons are excited from 1.2 eV below the valence band maximum (VBM) to 1.2 eV above the conduction band minimum (CBM). Asymmetric excitations are also considered, but the qualitative physics did not change, so they are not discussed here. The excitation density is ~$6.1 \times 10^{21}$ cm$^{-3}$, which is approximately 3.5% of the total valence electrons and can readily be achieved in experiments [38,51]. To study the effects of the excitation energy, we also consider LEE [the blue-colored regions in Fig. 2]. For a fair comparison, the set of ensembles used for statistical average is the same for GS, LEE, and HEE. Without the excitation, most of the atoms move only in the vicinity of their original octahedral positions, as shown in the insets of Fig. 2. This is because the lattice temperature here is well below $T_m$. On the other hand, with HEE, the positions of the atoms are significantly altered. Note that the distorted structures do not return to the initial octahedral structures in room-temperature MD simulations, which is indicative of a structural transition. In addition, the pair correlation functions [29] exhibits a flattening of the peaks and dips with respect to GS MD, and a shift in the first dip position of 3.5 Å (for crystalline GST) to a larger value of 3.8 Å (for amorphous GST).

To be more quantitative, Fig. 3 shows the time evolution of (ensemble-averaged) number of wrong bonds (WBs), fraction of distorted cations (DCs), and lattice temperature. The WB and DC are given as follows: We first define a critical radius for the nearest-neighbor distance $r_c$ = 3.5 Å, which is the first minimum in the pair correlation function. The coordination number of a cation is given by the number of neighboring anions within $r_c$. A cation is regarded as a DC when its coordination number deviates from that of ideal rocksalt (= 6). The WB is formed when a cation-cation or anion-anion bond length is shorter than $r_c$ [52,53]. In the crystalline phase, all cations (both Ge and Sb) are six-fold coordinated with Te. As the atoms undergo large lattice distortions, however, both WB and DC increase.

Figures 3(a) and (b) show that without the excitation, thermal motion of the atoms yields only a small number of WBs (< 10 per supercell) and DCs (< 19%). With excitation, significant increases in WBs and DCs are observed, especially for HEE. These results imply that HEE has undergone a more significant structural change than LEE. A more striking result is that all these changes occur when the lattice temperature in HEE remains at a level well below $T_m$, while in LEE the temperature increases significantly [see Fig. 3(c)]. In fact, the temperature in HEE is similar to that in GS, with fluctuations on the same order of magnitude. The results thus show that non-thermal phase transition only happens in HEE, while thermal activation affects the structure change in LEE considerably. Note that, although experiments clearly confirmed non-thermal phase transition in HEE [16,18], theoretical prediction in Ref. [26] is opposite. Our results show that the reason for the inconsistency may be rooted in the improper treatment of the carrier dynamics, namely, the adiabatic quasi-equilibrium distribution of the carriers, which does not include the effects of the non-adiabatic dynamics of the carriers, as described earlier.

To understand the excitation energy-dependent phase transition, let us revisit the carrier relaxation in Fig. 1. For simplicity, we discuss only the case of excited electrons, because the same applies to holes. While the carrier-lattice and carrier-carrier scatterings are the two dominant processes, the occupation of the electronic states has a significant effect on the carrier dynamics [29]. If the excited electrons mainly occupy high-energy states, the major process is carrier *relaxation* to lower-energy levels inside the conduction band. Here, CC scattering is more efficient than CL scattering [27,28] because of the large mass difference between electrons and atoms. Thus, for HEE, there is only negligible energy transfer to the lattice and hence there is no PES lowering, as shown in Fig. 1(b). On the other hand, if the excited electrons mainly occupy states near the band edges, the major process is electron *recombination* with holes. Here, a recombination by carrier-carrier scattering, i.e., the Auger process, inevitably increases the energy of the carriers, which is against the overall trend of carrier equilibration and is hence unlikely. Thus, the dominant process is phonon-mediated recombination [Fig. 1(c)] and the PES is significantly relaxed [Fig. 1(a)].

The analysis above is in line with our TDDFT-MD results. Figure 4 shows the excited carrier density $\rho$, obtained from its occupation [29], as a function of time. In the LEE, $\rho$ decreases monotonically with time, which implies that carrier recombination dominates. The significant increase in the lattice temperature shown in Fig. 3(c) indicates that the excited energy is dissipated mainly as heat. In contrast, in the HEE, $\rho$ increases at the beginning. Because Fig. 3(c) shows no significant temperature increase, we conclude that for HEE, CM by carrier-carrier scattering dominates. At 0.6 ps, $\rho$ reaches a maximum and after that it changes only slightly. This is a sign that the electron system reaches its own equilibrium, which is corroborated by the establishment of Fermi–Dirac distribution [29]. Note that previous TDDFT study also revealed the electron thermalization leading to a Fermi–Dirac distribution for the excited carriers [54]. The degeneracy at the Fermi level may lead to an "occupation gap", which, however, does not exist here due to the lack of degenerate states.

The PA model has been widely accepted as the mechanism for non-thermal phase transition in which a sizable $\rho$ is required [18,20-24]. Our simulation of GST, however, produces a qualitatively different picture in which not only a much broader energy range is covered but also it reveals a turnover from thermally-activated process to non-thermal process when the magnitude of $\rho$ has not changed. The lack of knowledge on the turnover in the past may be in part because the lack of an adequate method to approach the problem, such as the TDDFT, and in part because the focus has been on materials with relatively large band gap, for which carrier recombination and multiplication are unlikely at sub-ps time scale. For materials with smaller band gap such as GST and metals, the non-adiabatic dynamics is a strong function of the excitation energy. If the energy is as low as in LEE, one cannot even talk about non-thermal phase transition despite $\rho$ is large. Indeed, Fig. 3 (sky-blue line) shows that, when increasing $\rho$ for band-edge excitation to the level of HEE, only rapid thermal activation of the lattice takes place [55].

Besides GST, in the past InSb has been extensively studied, e.g., by using ultrafast time-resolved X-ray diffraction [15-18]. Based on the Debye–Waller model and the assumption of a static PES, the experimental results were interpreted as a carrier-density-dependent ionic motion, such as inertial dynamics and accelerated atomic disordering at a very-high $\rho$ [18], in line with the PA model. At a pump laser frequency centered at 800 nm (1.55 eV), the inertial dynamics and separately, the accelerated atomic disordering, may be interpreted as originated from the carrier-carrier scattering and a strong CM, respectively. Because InSb band gap of 0.17 eV is significantly smaller than the GST gap of 0.5 eV, InSb is also ideal for studying the turnover from thermally-activated to non-thermal processes.

Our results not only provide the mechanism for sub-ps non-thermal phase transition in GST, but also shed lights on carrier-induced phase transition at longer time scale for practical uses. To this end, we performed DFT-MD simulations, starting with final atomic structures and velocities of TDDFT-MD at 1 ps. To include the effect of excited carriers, we follow Refs. [22-26] to assume that the excited electrons and holes have reached a quasi-equilibrium that can be described by a Fermi–Dirac distribution. Different from those references, however, here we determine the electron temperature to be 5,500 K for HEE and 3,000 K for LEE from TDDFT-MD [29]. If the system already reached a global steady state, this is expected to have negligible effect on the follow-up simulation. Otherwise, one may expect a sharp peak in the ionic temperature at the onset of the DFT-MD simulation due to the discontinuity between the TDDFT and DFT PESs. Figure 3(f) shows pronounced discontinuities, particularly for HEE, as a result of the differences in the time-evolved TDDFT and static DFT wave functions.

The significant differences in ionic temperature, e.g., 1,200 K for HEE, corroborate with the fact that the system undergoes further structural distortions in the next 4 ps simulated by DFT-MD. Since for HEE, phase transition has taken place non-thermally in the sub-ps time scale, one may wonder if a prolonged heating is necessary. If one could remove the excess heat before the ionic temperature rises, for example, by an efficient heat transfer from nano-sized GST particles embedded in matrices or on a substrate, one should be able to realize non-thermal phase transition without actually raising the system temperature. In this regard, previous experiments have demonstrated the rapid non-thermal transition, in the absence of slow thermal transition, in 10- and 20-nm sized GST [56,57].

Finally, we should discuss the approximations we used and their validity. Regarding the memory effects, previous study has suggested [58] to measure them using the time derivative of the non-interacting Kohn-Sham kinetic energy. Using this approach, we find that the memory effects in our system are negligibly small [29]. An underestimated PBE band gap will affect the calculated excitation energy for the LEE. Correction to the band gap is expected to give rise to more energy transfer to ionic motion in the LEE, but such a quantitative difference should not alter the qualitative conclusions. In general, the results of CC scattering depend on the functional used in the calculation but the fundamental physics should not change. As a check, we performed LDA calculation and found that all the key observations in PBE hold. There are some concerns with the Ehrenfest approximation such as the lack of spontaneous phonon decay for carrier relaxation and the use of averaged potential energy surface [59,60]. We have estimated that the effects are either negligibly small or should not alter the qualitative results as these results reflect the fundamental physics of the carrier dynamics [29].

In summary, first-principles TDDFT-MD study reveals a new mechanism for non-thermal phase transition induced by CM. It provides a unified framework to understand the phenomena in a wide-range of materials. Furthermore, calculation suggests an unexpected turnover between thermally-activated and non-thermal processes as a function of the excitation energy, to be tested by experiment. The simulation further suggests that with high-excitation energy, non-thermal phase transition in GST can take place at a temperature significantly below $T_m$. Ways to take advantage of such a non-thermal phase transition in PCM-based non-volatile memory without significantly increasing the temperature and energy consumption are also considered.

JB was supported by the Basic Science Research Program through the National Research Foundation of Korea (NRF) funded by the Ministry of Science, ICT & Future Planning (NRF-2015R1C1A1A02037024). SBZ was supported by the U.S. Department of Energy (DOE) under Grant No. DE-SC0002623. YYS acknowledges the support of the National Science Foundation under Award No. CBET-1510948. FG was supported by the Nuclear Regulatory Commission (NRC), USA. Research was performed at the Center for Computational Innovations at RPI and using a national scientific user facility in the Environmental Molecular Sciences Laboratory, sponsored by the DOE's Office of Biological and Environmental Research.

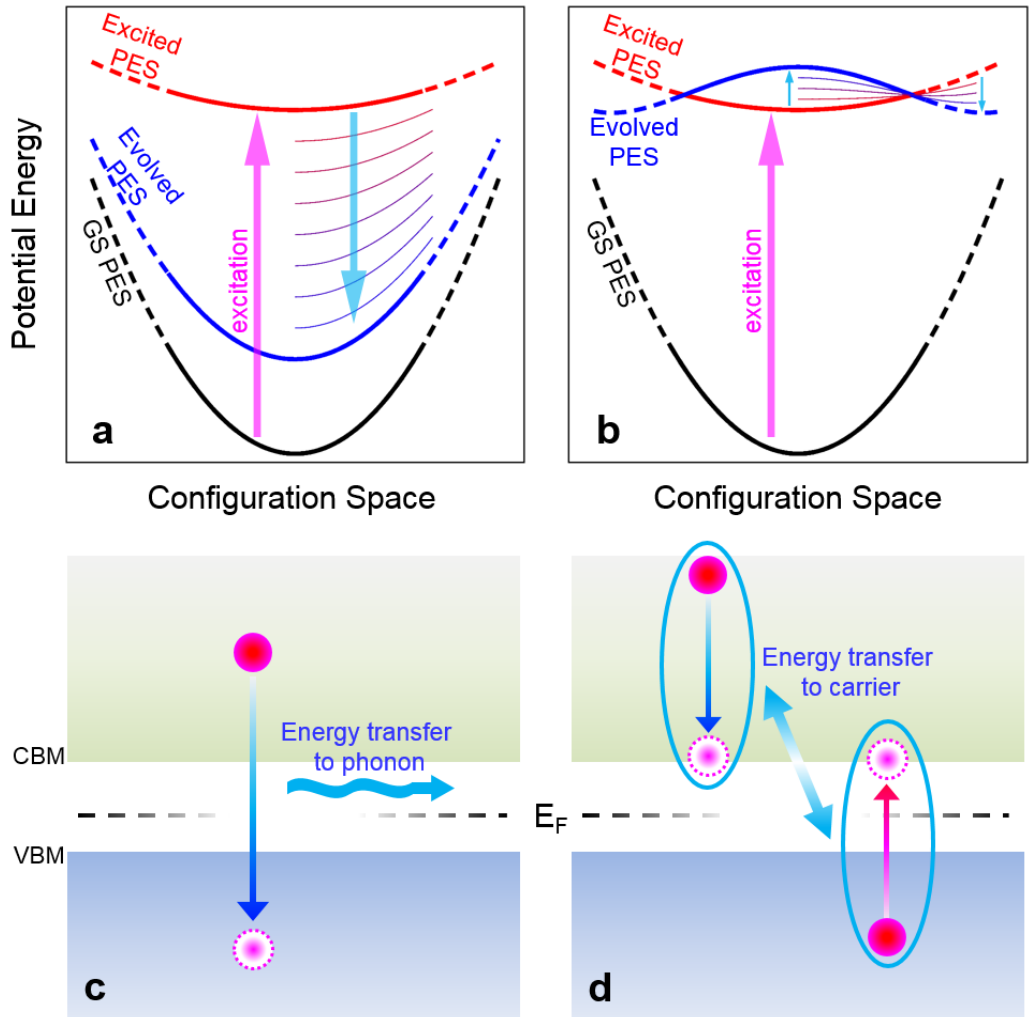

**FIG. 1.** (Color online) Schematics of PES evolutions by (a) CL and (b) CC scatterings. A sufficiently large amount of excitation leads to shallower PES. In (a), since the energy of the excited carriers is transferred to lattice, its PES is vertically dropped. In (b), the excited carrier energy is exchanged within the electronic system, so PES moves horizontally. The CL and CC scattering processes are shown in (c) and (d), respectively. In (c), an electron recombines with a hole in the valence band, reducing ρ, and in (d), the electron relaxation increases ρ, which is referred to as CM.

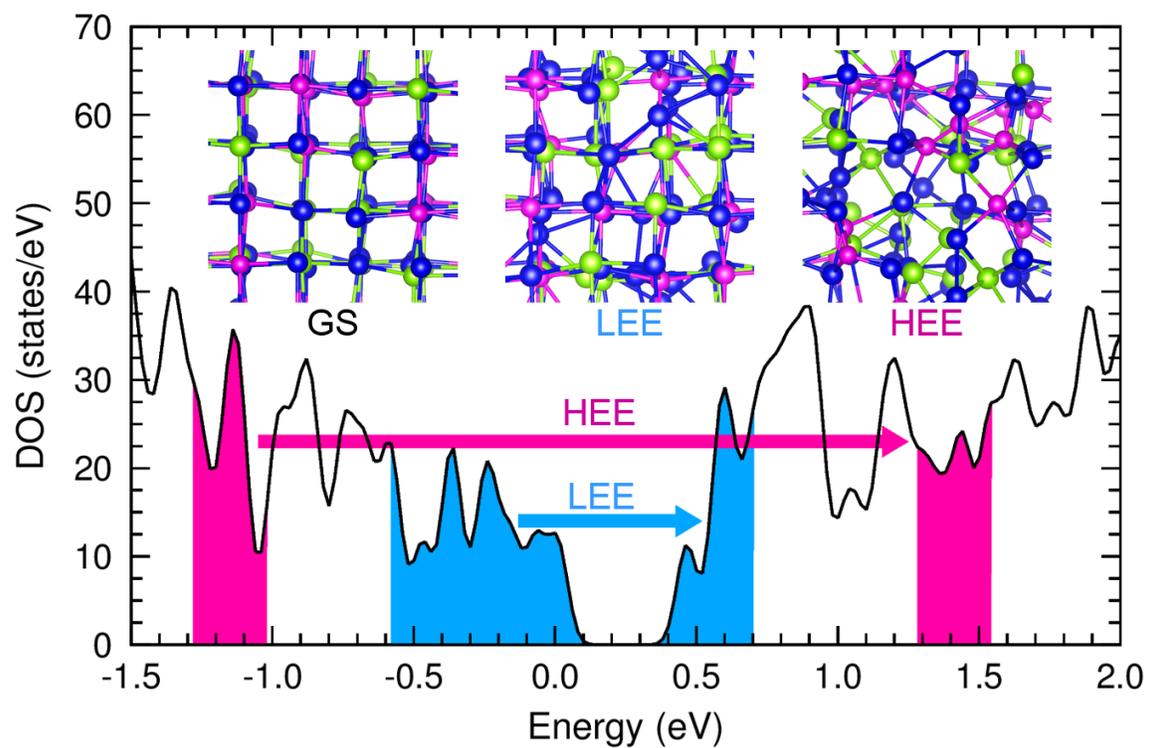

**FIG. 2.** (Color online) Density of states (DOS) of GST. Blue-to-blue is for LEE and pink-to-pink is for HEE. Insets show atomic structures at the end of GS, LEE, and HEE MD simulations. Pink, green, and blue balls are Ge, Sb, and Te atoms, respectively.

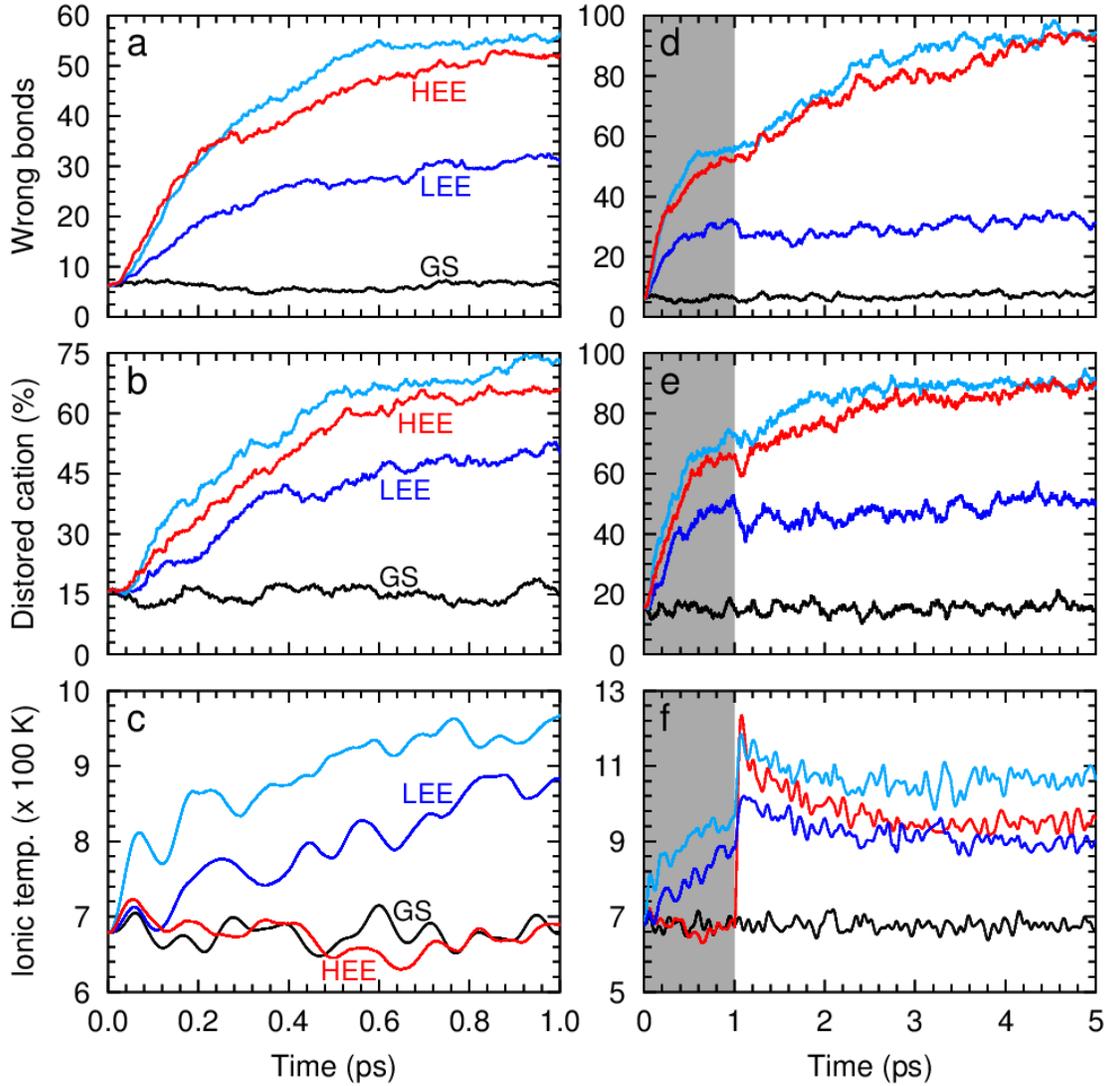

**FIG. 3.** (Color online) Time evolution of (a) number of wrong bonds, (b) percentage of distorted cations, and (c) ionic temperature in TDDFT-MD. Black, blue, red, and sky blue lines correspond to GS, LEE, HEE, and high-carrier-density LEE, respectively. Standard DFT-MD simulations from the final step of TDDFT-MD (see the main text) are shown in (d), (e), and (f). The gray regions from 0 to 1 ps are the TDDFT-MD results in (a), (b), and (c).

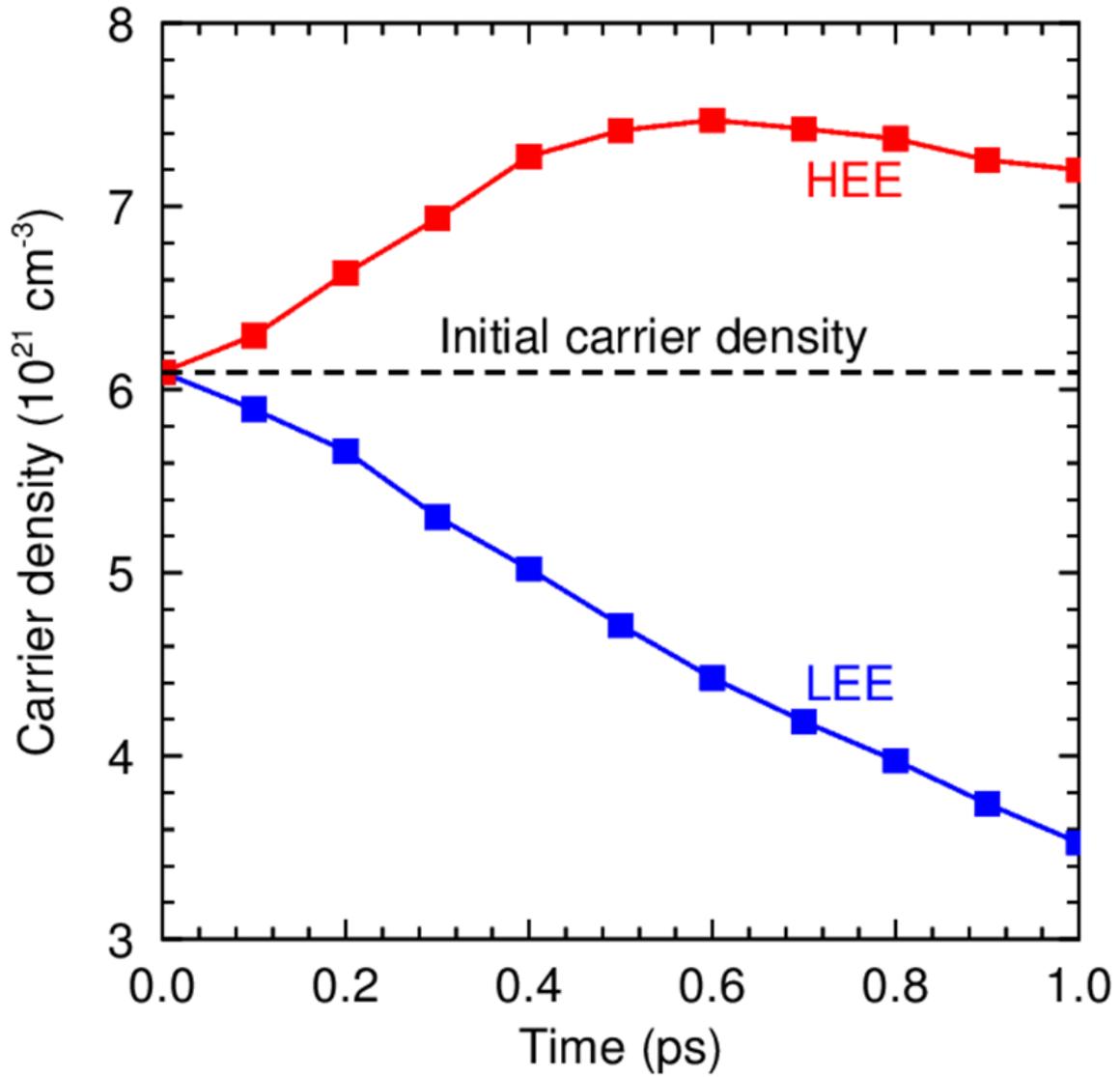

**FIG. 4.** (Color online) Time evolution of the excited electron density for LEE (blue) and HEE (red). Note that the excited electron and hole densities are the same.